\documentstyle[12pt,aaspp4]{article}
\singlespace
\def\kms{km\thinspace s$^{-1}$\ }
\def\ms{m\thinspace s$^{-1}$\ }
\def\vsini{$v${\thinspace}sin{\thinspace}$i$}

\slugcomment{{\it Ap. J.}, in press}

\lefthead{Hatzes and Cochran }
\righthead{Does $\tau$ Bootis Have  a Planetary Companion?}

\begin{document}

\title{A Search for Variability in the Spectral Line Shapes of
$\tau$ Bootis: Does this Star Really Have a Planet?}

\author{Artie P. Hatzes and William D. Cochran}
\affil{McDonald Observatory, The University of Texas at Austin,
    Austin, TX 78712}

\setcounter{figure}{0}
\begin{abstract}
An analysis is made of the spectral line shapes of $\tau$ Bootis 
using high resolution (0.026 {\AA}) and high signal-to-noise ($S/N$ 
$\approx$ 400) data in an
effort to confirm the planet hypothesis for this star. Changes
in the line shape are quantified using  spectral line bisectors
and line residuals. We detect no variations in either of these
quantities above the level of the noise in the  data. One spectral line,
Fe I 6213 {\AA}, does show a hint of sinusoidal variations in the bisector
velocity span when phased to the radial velocity period of 3.3 days,
but this is not seen in the bisectors for two other lines, nor in the
line residuals. Comparisons of the data to the bisector and residual
variations expected for nonradial pulsations indicate that we can exclude
those sectoral nonradial modes having 
$m$ $>2$ and all sectoral modes with $k$ $>$ 1, 
where $k$ is the ratio of the horizontal
to vertical velocities for the pulsations. The lack of line shape
variability and the 469 {\ms} radial velocity amplitude is still consistent
with nonradial sectoral modes $m$ = 1, and possibly $m$ = 2, but with 
$k$ $\approx$ 1, which is at least 3 orders of magnitude less than the
predicted value given the 3.3 day period of $\tau$ Bootis. Such low values
of $k$ can probably be excluded given the lack of photometric variations
for this star. Although the measurements presented
here do not prove, without any doubt,
that $\tau$ Boo has a planetary companion, they do add significantly
to the increasing body of 
evidence in favor of this hypothesis.
\end{abstract}

\section{Introduction}

	The exciting discovery of the planetary companion to
51 Peg (Mayor \& Queloz 1995) marked a breakthrough in the
search for extra-solar planets.
After years of unfruitful searches, precise radial velocity (RV)
surveys have uncovered several extra-solar planets since the 
51 Peg discovery (Butler et al. 1997;
Cochran et al. 1997; Noyes et al. 1997; Butler \& Marcy 1996).
One of the more unexpected
results of these surveys is the fact that giant, Jovian-mass
planets can exist in very short-period orbits around solar-type stars.
Although first received with skepticism (primarily
because it was so unexpected), planet formation theories can
now provide a natural explanation for the occurrence of these peculiar
systems (Boss 1995; Lissauer 1995; Lin et al. 1996).
These systems, termed the ``51-Peg systems" by  Butler et al. (1997)
are characterized by planets with  masses in the range of
0.5-4 $M_{Jupiter}$ and orbital periods of 3.3--14 days (0.05 -- 0.1 AU).
Members of this class include 51 Peg, $\tau$ Boo, $\rho^1$ CNc, 
and $\upsilon$ And.

	Because RV  measurements provide only an indirect
means of detecting a planetary companion, one can never be absolutely certain
that the RV signal is due to the reflex motion of the star.
After all, pulsations, both radial and nonradial, as well as stellar
surface structure can also produce  RV variations which
would mimic the signal caused by  a planetary companion. It is
for these reasons that ancillary measurements are needed to confirm
all planetary discoveries made by RV measurements. Photometric 
measurements of many of the 51 Peg-like stars have established
a constant light level to a limit of 0.4 mmag (Baliunas 
et al. 1997). These results provide strong
support to the hypothesis that these are  indeed planetary
companions to the 51 Peg-like stars.

	Recent work, however, has questioned whether  a planet
actually exists around the prototypical
system, 51 Pegasi. In a controversial
paper, Gray (1997) reported changes in the spectral line shapes 
(as measured by the spectral line bisector)
of the Fe I 6253 {\AA} line with the same period as the RV variations
in this star.
Gray \& Hatzes (1997) showed that both the changes in 
the line shapes and the RV variations
for 51 Peg could be modeled by nonradial pulsations without
the need to invoke a planet. Furthermore, the nonradial mode 
would have to be a g-mode oscillation which means that most of the
atmospheric motion is in the horizontal direction. 
 Since these
produce small distortions in the projected area of the star,
the expected photometric amplitude would
be well below detection levels.
However,
the spectral variability for 51 Peg was recently refuted
by Hatzes, Cochran, \& Bakker (1998) using higher quality 
data; this star most likely has a planet. 
Even though the dispute over the planet of  51 Pegasi has 
been resolved, it is still worthwhile
to confirm the other short-period extra-solar planet discoveries 
and a
detailed examination of
the spectral line shapes may provide the most
stringent test for such a confirmation.

	With a semi-amplitude of 469 {\ms} and a period
of 3.3 days the ``planet'' around
$\tau$ Boo is the  most massive  of the 51-Peg  systems
($m$ sin $i$ = 3.7 $M_{Jup}$). Ca II H and K measurements
imply a rotation period of 3.4 -- 4 days  for the central
star (Baliunas, Sokolof, \& Soon
1996; Baliunas et al. 1997), so there is some concern that the
RV period may be due to rotational modulation. However,
the low upper limit ($<$ 1 mmag) to any photometric variability of
this star argues against surface structure and possibly pulsations
as a cause for the RV variations (Baliunas et al. 1997). There
should also be a concern that this star is a nonradial pulsator.
Baade \& Kjeldsen (1997) found evidence for line profile
variations in the star $\epsilon^2$ Ara that were indicative
of an m $\sim$ 4 mode. The spectral type of $\epsilon^2$ Ara, F6V, is
uncomfortably close to that of $\tau$ Boo (F7V). The RV amplitude
for $\tau$ Boo, however,  is so large that if it was  due to nonradial
pulsations  then changes in 
spectral line shapes would be large and readily measurable. 
In this paper we 
analyze the spectral line shapes of $\tau$ Boo 
in an effort to confirm or deny the planet hypothesis for this star.

\section{Data Acquisition}

	Data were acquired using the 2-D coud{\'e} spectrograph
(Tull et al. 1995)
of the McDonald Observatory's 2.7-m  Harlan J. Smith Telescope. 
This  instrument is
a cross-dispersed echelle spectrometer used with a Tektronix 2048$\times$2048
CCD detector (24 $\mu$m pixels) and it can be operated in a ``low'' resolution
($R$ $\sim$ 60,000)
focus with nearly complete wavelength  coverage in the 
range 4000 {\AA} -- 1 $\mu$m,
or in a ``high''  resolution focus with more limited and non-contiguous
 wavelength coverage ($\approx$ 400 {\AA}).
The detector was placed at the  high resolution (F1) focus for this
study.
A projected slit width of 2.4  pixels at the detector
resulted in a working resolving power of 240,000 (resolution = 0.026 {\AA}).

	Shortly after commissioning of the 2-d coud{\'e} we noticed
that when  using instrument in its high resolution mode the
theoretical  resolving power could not be achieved. 
Furthermore, this limiting
resolution seemed to degrade with time.
Tests with a pinhole aperture produced two images at the camera focus.
Most of the time  these images fell on a line perpendicular to the
dispersion which should not affect the instrumental profile (IP)
when co-adding all CCD rows contained in a spectral order. However,
there were instances when these two images were not aligned precisely
perpendicular to the dispersion, and this would degrade the IP.
Further tests showed that by masking the top half of the spectrograph
collimator
we could isolate the one pinhole image which remained fixed with time.
This also enabled us to recover the full resolution of 
the spectrograph.
Work is in progress to
isolate and rectify the problem (it is suspected
that this effect is due 
to stresses imposed on the cross-dispersing
prisms in their mount).   
All observations were made with the top half of the collimator masked.
This is an investigation of subtle changes in the spectral line shapes
which could be affected by changes in the IP.
Consequently, 
we chose to sacrifice photons at the expense of a stable
IP. 
An examination of the thorium emission lines taken on the nights showed
that the rms scatter of the full width to half maximum (FWHM) varied by no
more than 10\% (0.002 {\AA}) from observing run to run. 
Considering that the  FWHM of the spectral
lines of $\tau$ Boo is about 0.5 {\AA} such a small
IP variability should not affect the spectral
line shapes.

	Table 1 lists the journal of observations which include
Julian day of the mid-exposure, exposure time, and RV phase.
Phases were computed according to the orbital ephemeris of 
Butler et al. (1997):
\begin{equation}
T_{max} = 2,450,235.41  + 3.3128 E
\end{equation}
where $T_{max}$ represents the phase of maximum radial velocity.

\section{Results}
\subsection{Spectral Line Bisectors}

	Spectral line bisectors, which consist of the locus
of midpoints to horizontal line segments extending across the profile,
provide sensitive measures of changes in the line shapes caused
both by surface features (Toner \& Gray 1988) or nonradial pulsations
(Hatzes 1996; Gray \& Hatzes 1997). However, these types
of measurements are
less effective for stars with rapid rotation since the decreased
slope and depths of the lines greatly increase the error of the bisector
measurement. Also, rapid rotation increases the likelihood
for line blending. With a rotational velocity of 15 {\kms} $\tau$ Boo is a 
less than ideal candidate for bisector measurements. Even so, an attempt
was made to search for changes in the spectral line shapes using line
bisectors. 

	For a spectral line to be a suitable candidate for bisector measurements
it must be relatively strong and blend-free. The rapid rotation of
$\tau$ Boo obviously 
works against these criteria. In the spectral region covered
by our observations there were only three lines suitable for bisector
measurements. These are listed in Table 2.

\begin{figure}
\plotone{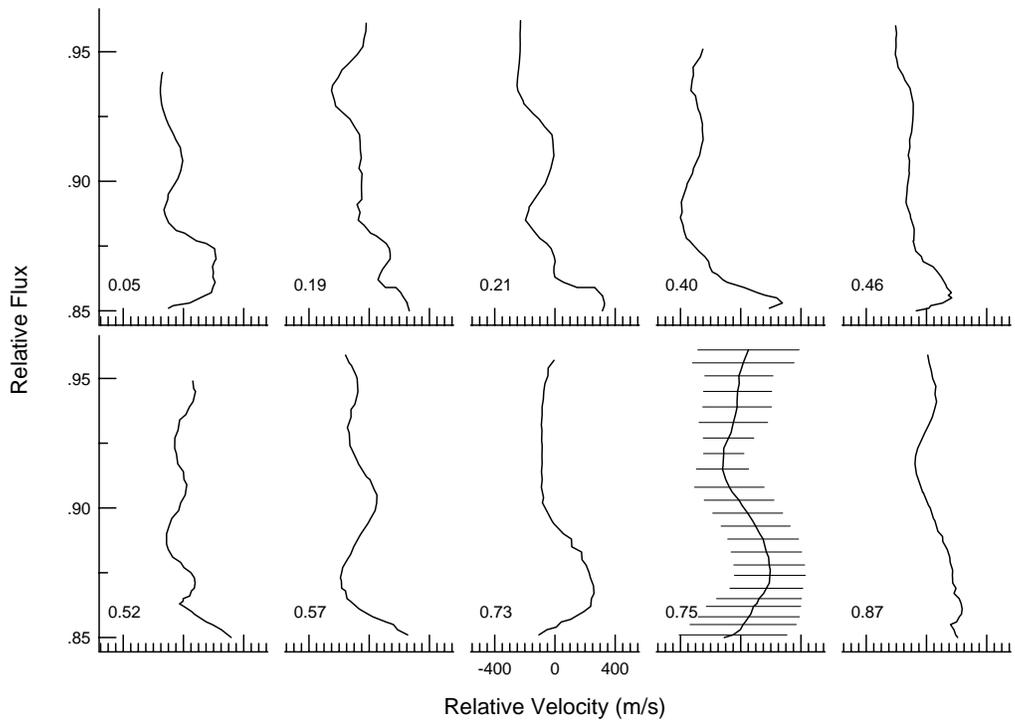}
\caption{The spectral line bisectors as a function of phase
for Fe I 6213. The error bars shown at phase 0.75 represent
typical errors.}
\label{bisect1}
\end{figure}

\begin{figure}
\plotone{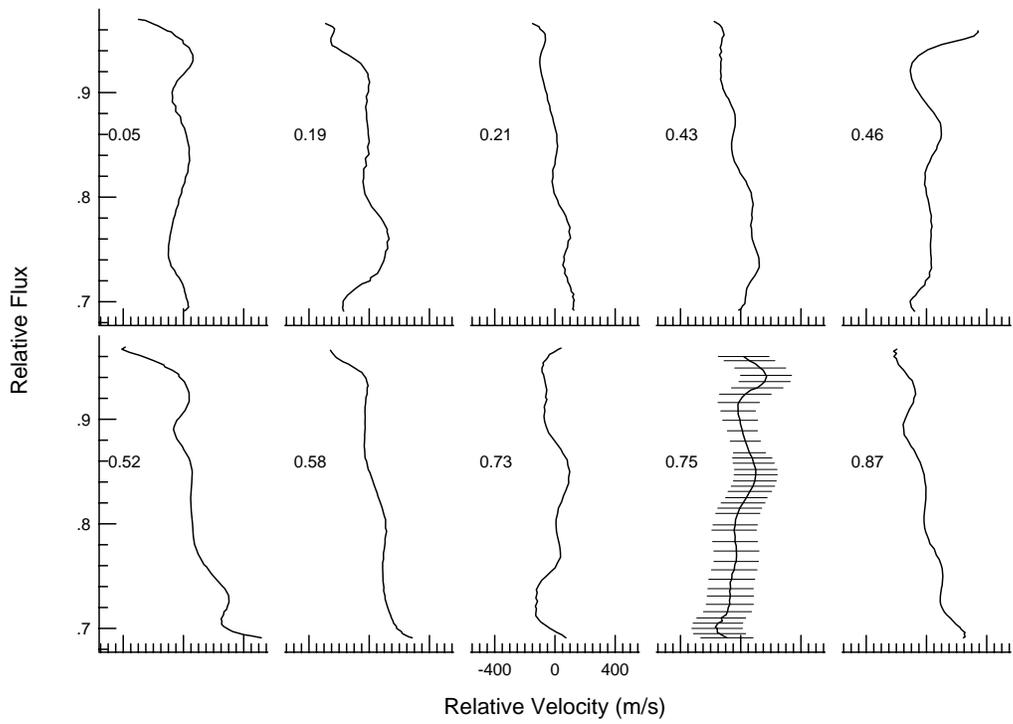}
\caption{The spectral line bisectors as a function of phase
for Ca I 6439. }
\label{bisect2}
\end{figure}

	Prior to computing the individual line bisector 
the spectra were
smoothed with a Gaussian having a full width at half maximum of
4 pixels. (Due to the broad lines in $\tau$ Boo we could afford to increase
the overall signal-to-noise ratio at the expense of degraded resolution.)
Individual bisectors (2--4) from a given 
night were co-added to produce
a nightly mean. Figures~\ref{bisect1} and ~\ref{bisect2} shows
the line bisector variations for the Fe I 6213 and Ca I 6439 {\AA}. (The
line bisectors of Fe I 6337 {\AA}  show comparable variations, or lack thereof,
and are not shown. The velocity span measurements for this line, 
however, will be used in
\S 3.3). Error bars are shown only for the bisector at phase 0.75
and these represent a typical  value for the measurements. These 
were estimated using the noise level as measured in the continuum
and the slope of the flux profiles (Gray 1988).
There 
appear to be no systematic variations with phase for the bisectors
that is above the noise level. In \S3.3 we quantify the possible
changes in the bisector slopes.

\subsection{Line Residuals}

	Because of the line broadening in $\tau$ Boo, spectral line
bisectors may not be the best way to search for spectral variability
in this star. The rapid rotation, on the other hand, may 
enable us to see  directly the distortions in the spectral line profiles
due to pulsations. It is well known that  for stars with rapid rotation
both nonradial pulsations and surface structure produce distortions
that propagate through the spectral line profile. These are best seen
in stars rotating at {\vsini} $>$ 25 {\kms}, but should still be visible
in spectral line profiles with the same broadening as $\tau$ Boo.
A convenient means of measuring the amplitude of these distortions
is through the line residuals (e.g. Kennelly et al. 1992; Baade \& 
Kjedsen 1997).

\begin{figure}
\plotone{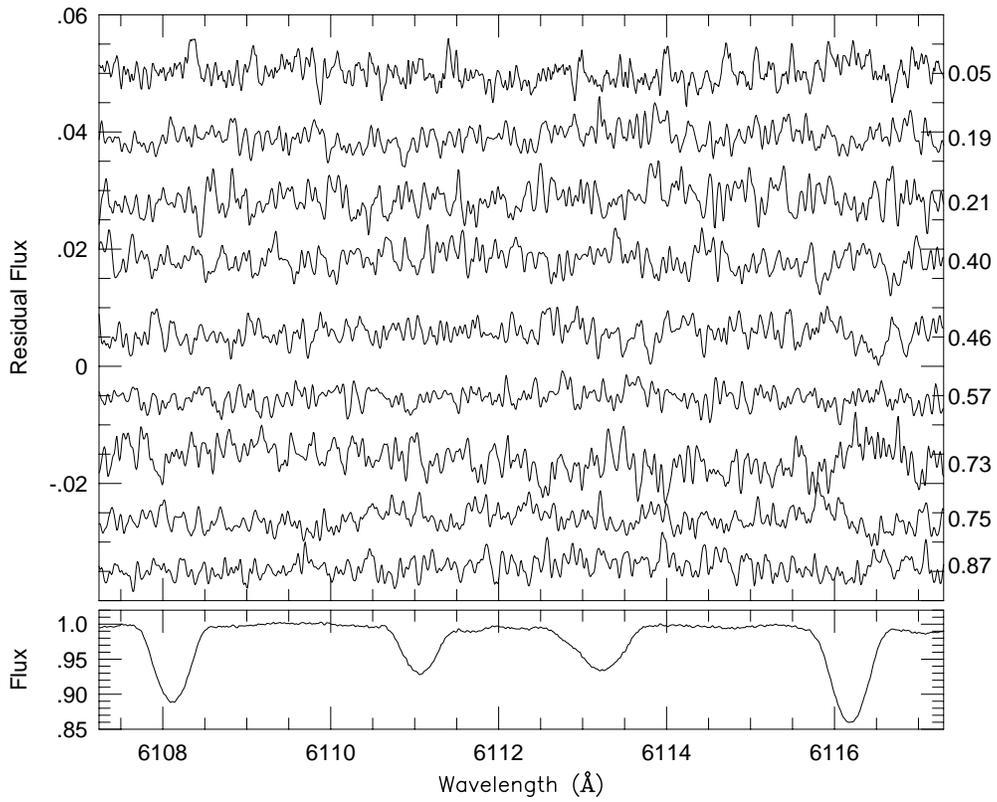}
\caption{(Top) The spectral line residuals as a function of phase
for  the 6107 -- 6117 {\AA} spectral region of $\tau$ Boo.
(Bottom) The grand mean spectrum in the same wavelength
interval.}
\label{resids1}
\end{figure}

\begin{figure}
\plotone{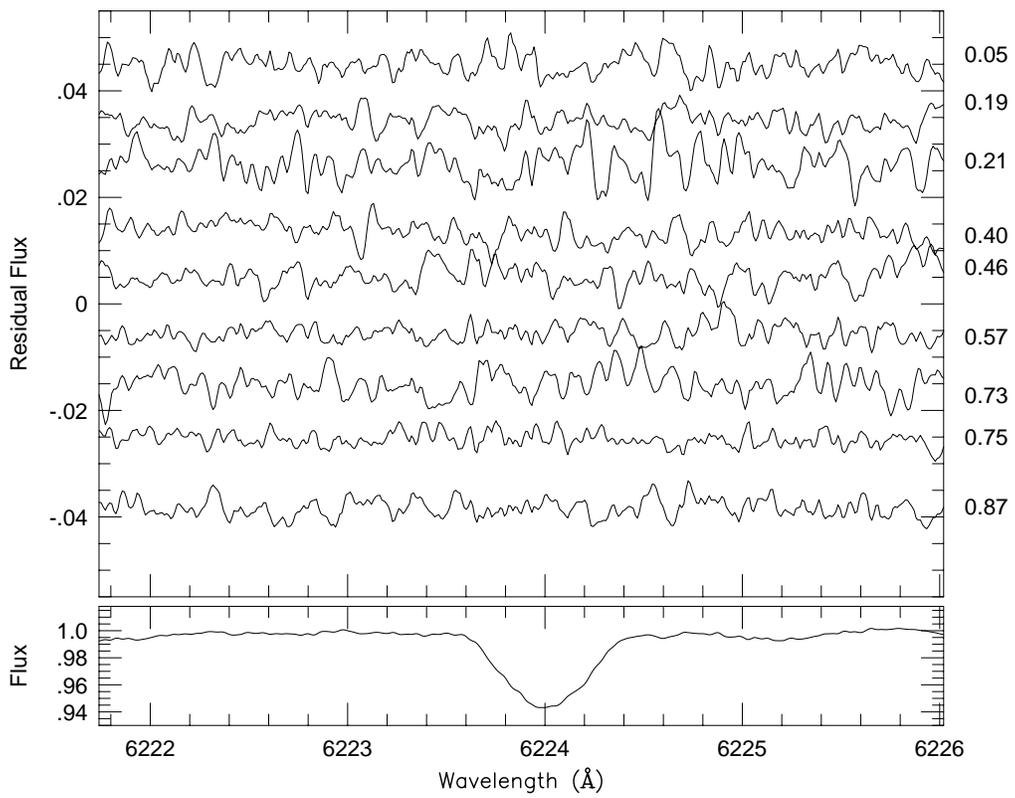}
\caption{The spectral line residuals for Fe I 6224 {\AA} (top)
and the grand mean spectrum (bottom).}
\label{resids2}
\end{figure}

	Line residuals were computed by first co-adding all observations
to produce a grand mean. Prior to adding the spectra they were
all aligned to the rest frame of the star. (Again, all observations
were smoothed by a 4-pixel Gaussian prior to computing
line residuals.) Observations on a given night were than combined to
produce a nightly mean.  (The signal-to-noise ratio of these nightly means
was $\approx$ 400 per pixel.)
The line residuals were taken as the difference between the nightly means
and the grand mean. Table 3 lists the spectral lines that were used 
to search for residual variations. Figure~\ref{resids1} and Figure~\ref{resids2}
are representative residuals for the four spectral lines in the
wavelength interval 6108 {\AA} -- 6117\,{\AA} and the Fe I line 6224 {\AA}.
There appears to be no phase-dependent distortions across the spectral
line profile and any distortions which may be present are at 
the level of the noise (as indicated by the scatter of
the residuals in the continuum).

\subsection{Limits on Nonradial Pulsations}
	
	At the present time, 
nonradial pulsations seems to be the only hypothesis,
other than a planetary companion,
that may provide an explanation	for the RV variability of $\tau$ Boo.
Even though the estimated rotational period for $\tau$ Boo is near that
of the RV period, surface features (spots, etc) can be excluded
as a possible hypothesis not only on the basis of the lack of 
photometric variations, but also from the lack of spectral distortions
as evidenced in our bisector and line residual measurements. It is
difficult for any  surface feature to account for the  large
RV amplitude in $\tau$ Boo
without producing measurable changes in the line shapes.

	It is possible, however, for nonradial pulsations to produce
significant RV variability without large variations in the spectral line
shapes (Hatzes 1996). The question is, whether these can produce
the large RV amplitude  seen in $\tau$ Boo without measurable
line distortions. In this section we 
place limits on any possible nonradial pulsation
mode which may be present on $\tau$ Boo. We considered all
sectoral nonradial modes that were 
capable of producing the observed RV amplitude. Furthermore,
the shapes of the RV curves from nonradial pulsations are  nearly
sinusoidal, consistent with the nearly zero orbital eccentricity.

	The expected line bisector and line residual variations
from nonradial pulsations were calculated using the prescription
outlined in Hatzes (1996). Nonradial pulsations can be described in
terms of spherical harmonics $Y_{\ell, m}$ where $\ell$ is the 
degree and $m$ the azimuthal order. 
Only sectoral modes ($\ell$ = $m$) 
were considered as these should produce the largest integrated RV amplitude.
The pulsational amplitude for each mode was chosen such that the integrated
RV amplitude from the  oscillations
was 469 {\ms}, i.e. the observed RV amplitude in $\tau$ Boo.

In calculating the local velocity field on the stellar
surface due to nonradial pulsations  one needs to know
the ratio of the horizontal
to vertical velocity amplitudes of the atmospheric motions.
Normally this is not a free parameter, but can be calculated
by
\begin{equation}
{k =} \left({Q} \over {0.116}\right)^2
\end{equation}
where $Q$ is the pulsation constant =  $\Pi({\rho}/{\rho_{\odot}})^{1/2}$.
($\Pi$ is the pulsation period in days, $\rho$ is the mean density of the
star, and $\rho_{\odot}$ is the mean density of the Sun).
For $\tau$ Boo the appropriate value of k is $\sim$ 1000.  However,
there is a well know ``k-parameter problem'' 
where in modeling the line
profile variations of known nonradial pulsators, the profiles can be
better fit with low values of $k$ ($\sim$ 0) even though the pulsation
period implies the need for much larger values of $k$ (Lee \& Saio 1990). 
Consequently,
$k$ was treated as a free parameter. 

A convenient means of quantifying a line bisector is through its velocity
span. Due to the large errors in the bisector data  we chose
to measure  this velocity span using a linear regression  of 
all bisector points within two arbitrarily chosen flux points.
As can be seen from  Figs. 1 and 2 measuring the span using only
two bisector points can result in 
grossly different velocity span
measurements depending on which  end points are chosen.  The slope of
the linear regression through all bisector points is less sensitive
to the choice of end points.
Comparisons of this span
to those predicted by nonradial pulsations were done in a consistent manner,
i.e. the bisector from a synthetic spectral line profile having nonradial
pulsations was computed and the velocity span was calculated 
using a linear regression through all bisector points
within the specified flux values used in the data bisectors.
The synthetic data, like the real data,
 were also smoothed by a Gaussian having a FWHM of
4 pixels.

\begin{figure}
\plotone{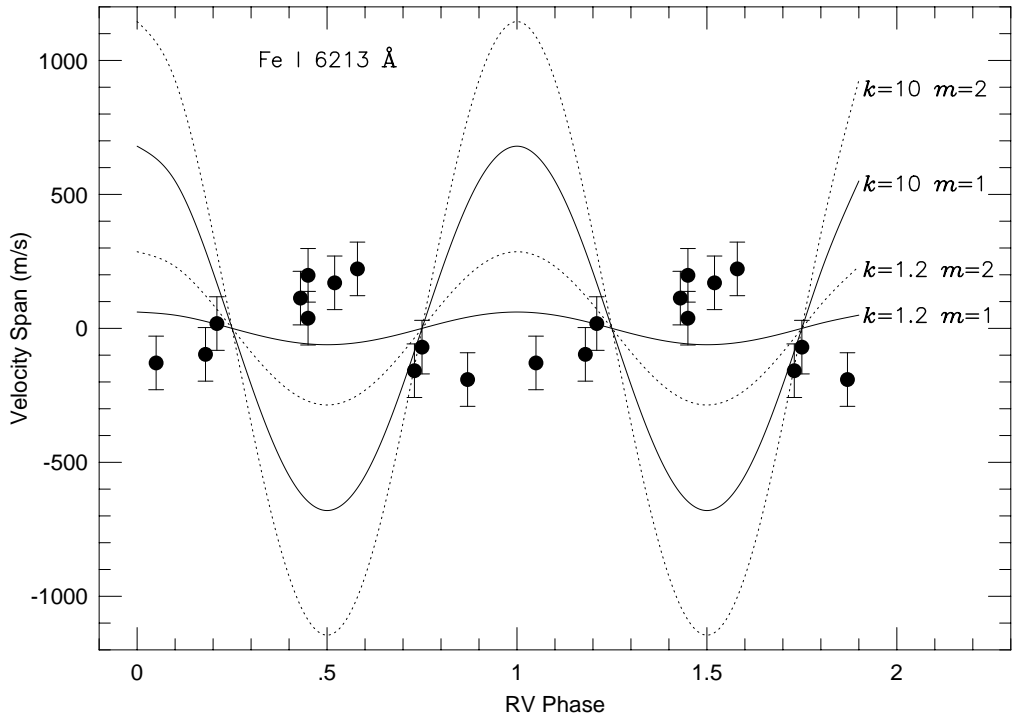}
\caption{Bisector velocity span for Fe I 6213 {\AA}. The lines
show the predicted variations for various pulsation modes.}
\label{span1} 
\end{figure}

\begin{figure}
\plotone{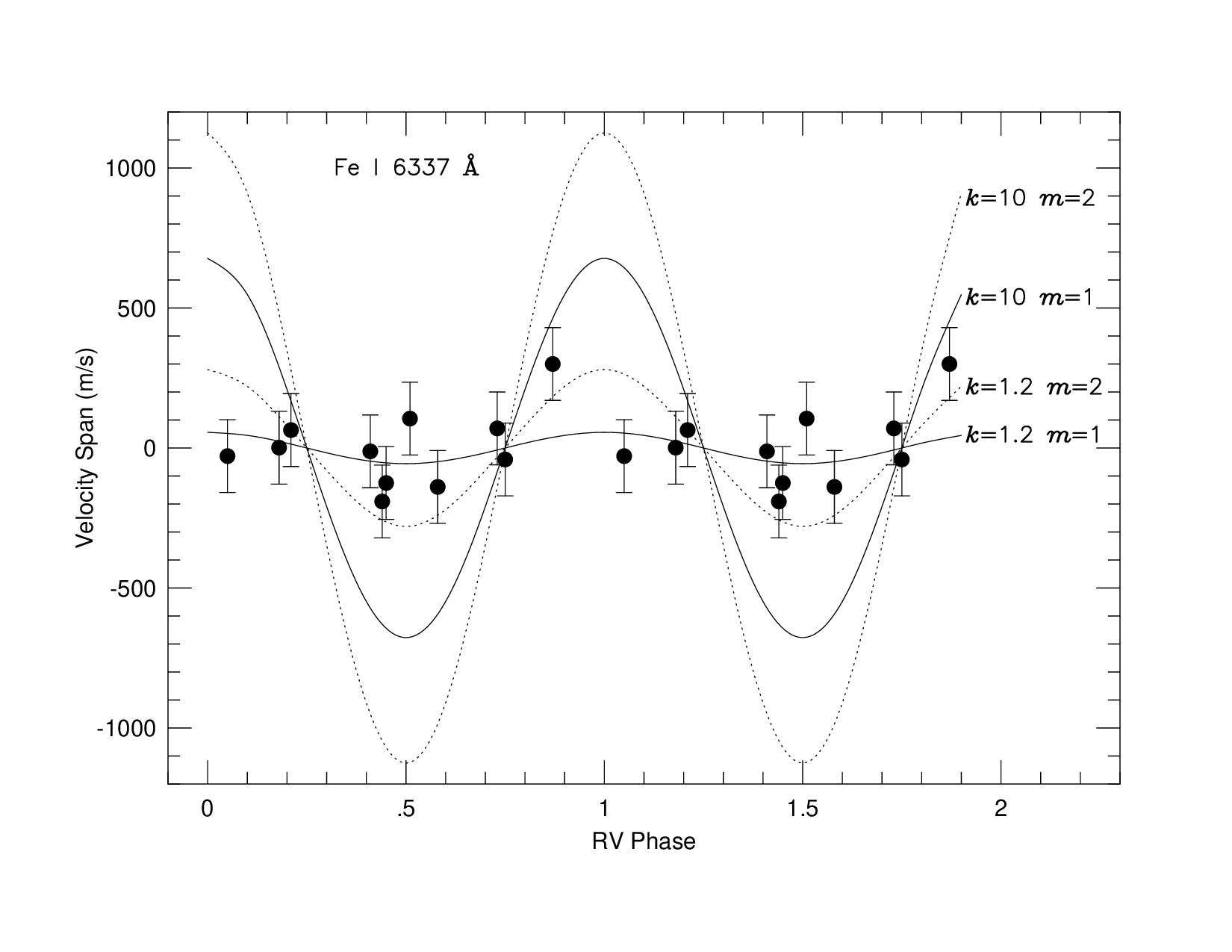}
\caption{Bisector velocity span for Fe I 6337 {\AA} compared
to the predicted variations from nonradial pulsations.}
\label{span2} 
\end{figure}

\begin{figure}
\plotone{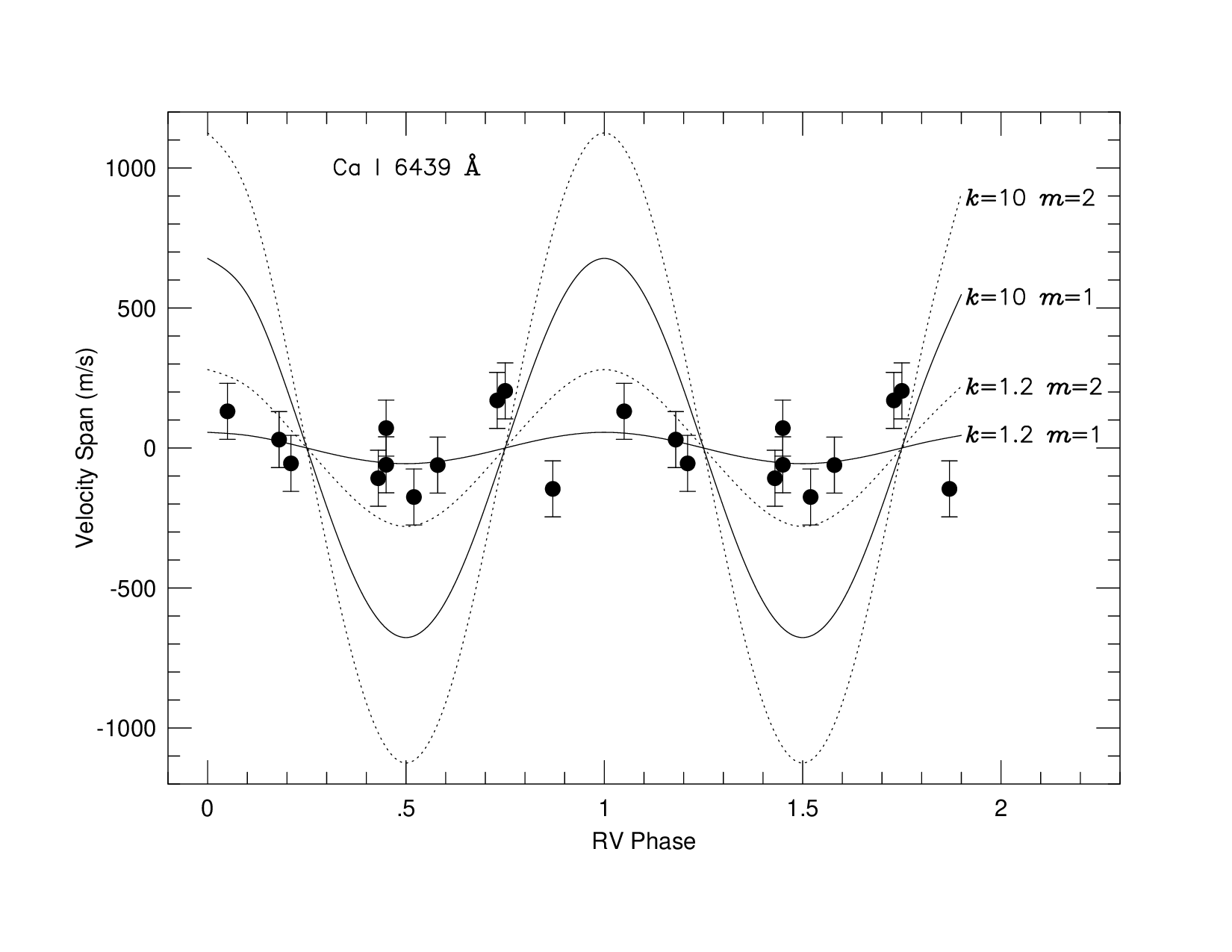}
\caption{Bisector velocity span for Ca I 6439 {\AA} compared
to the predicted variations from nonradial pulsations.}
\label{span3} 
\end{figure}

	Figures~\ref{span1} -- \ref{span3} show the bisector
velocity span measurements for the Fe~I~6213 {\AA}, Fe~I~6337 {\AA},
and the Ca~I~6439 {\AA} lines, respectively.  The averaged velocity span
measurements are tabulated in Table~\ref{spantable}.
The error was estimated by taking
the standard deviation, $\sigma$, of the individual span measurements clustered
around phase 0.45. This $\pm\sigma$ was plotted as the
error on each of the data points.
Superimposed on these figures are the predicted velocity span variations
from nonradial modes having $k$=10, $m$=1,2 and $k$=1.2 (a canonical
value for g-mode oscillations),  $m$=1,2.
Use of the predicted $k$ value ($k$ $\approx$ 1000) for $\tau$ Boo 
resulted in span variations
much larger than those shown in the figures. This was
also true for sectoral modes with $m$ $>$ 2, regardless of the
$k$-value. We can thus exclude those pulsation
modes with $m$ $>$ 2 and all modes with
$k$ $\gg$ 10. Only the $m$ = 1 with
$k$ = 1.2  would produce bisector span variations  which would be hidden
by the noise in our data.
The presence of an $m$ = 2, $k$ = 1.2 mode is marginally possible
and, in fact, the  Fe I 6213 {\AA} bisector spans do show a hint
of sinusoidal variations with an amplitude  comparable to 
that expected for an $m$ = 2 (albeit out-of-phase). However, we deem this to
be merely noise as there are no obvious sinusoidal variations in the span
measurements for the other two lines.

	Even though the phased bisector velocity  span measurements
show, for the most part, no obvious sinusoidal variations,
a periodogram analysis was still performed these measurements
(Table~\ref{spantable}). There was no statistically significant
power at the RV period of 3.31 days in any
of the periodograms. The false alarm probability for any power at the
RV period was 0.4 for the Ca I 6439 {\AA} line and
0.2 for the Fe I 6337 {\AA} line. Power did appear near the appropriate
frequency in the periodogram
for the Fe I 6213 {\AA} span measurements with a moderately low
false alarm probability of 0.02. Again, we do not deem this significant
based on the lack of power at the appropriate period
in the other periodograms. (We
should point out that the best line for bisector measurements, the
Ca I 6439 {\AA}, showed the highest false alarm probability.)

\begin{figure}
\plotone{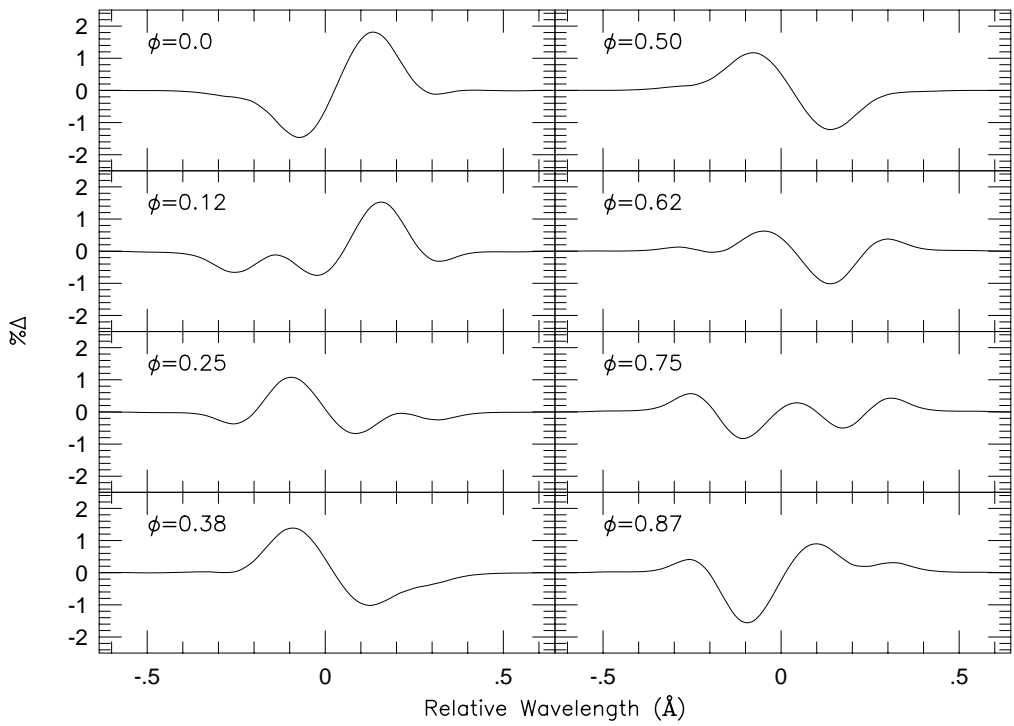}
\caption{Predicted spectral line residuals as function of 
phase for an $m$=2 nonradial mode with $k$ = 10.
}
\label{nrpres} 
\end{figure}

\begin{figure}
\plotone{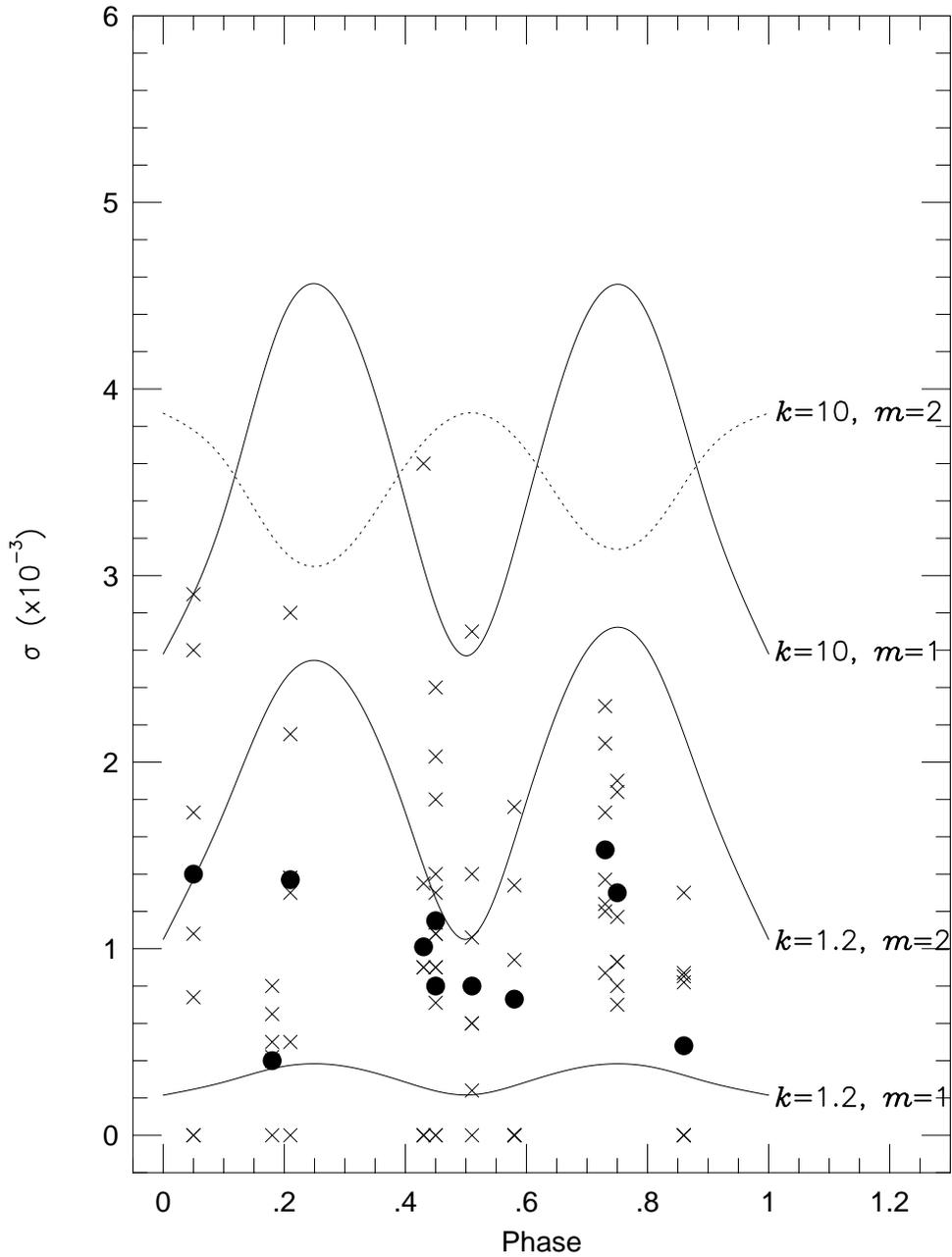} 
\caption{The crosses represent the standard deviation,
$\sigma$,  of the line residuals
across the spectral line profile for the seven spectral features
that were examined. The solid points represent the average of all
seven lines. The lines represent the predicted $\sigma$ as a function 
of phase for various nonradial modes.}
\label{sigma} 
\end{figure}

	Nonradial pulsations should also be visible in the spectral line
residuals. Figure~\ref{nrpres} shows the percent change
in the model spectral line residuals 
for an $m$=2, $k$=10 mode. The distortions have an amplitude
of a few percent, much larger than the observed distortions in 
Figure~\ref{resids1} and \ref{resids2}. (The amplitude of the distortions for
the $m$ = 1, $k$ = 1.2 mode are about 0.5\%.) Figure~\ref{sigma}
shows the standard deviation of the predicted spectral line residuals for 
various pulsation modes as a function of RV
phase. The standard deviation was computed over the
full width of the spectral line.  The crosses represent the standard deviation
of the line residuals of the actual data, again computed over the full
width of the spectral line. Individual measurements are shown as crosses
while the solid dots represent a mean value for all seven spectral lines
that were examined.
Part (if not all) of the rms scatter that was computed for the real
data 
may be due to photon noise. In order to correct for this, the $\sigma$ over
an equivalent wavelength region in the continuum was calculated and
subtracted in quadrature from the $\sigma$ computed across the 
spectral line. 

	It is not strictly correct to subtract the $\sigma$ in
the continuum since the signal-to-noise ratio in the spectral
line decreases from the continuum value. For line depths
of 10--15\% and $S/N$ $\approx$ 150, this effect should
result in a value $\sigma$ $\sim$ 0.0015 for the residuals
across the line profile. This is comparable to the mean level
of $\sigma$ for the line residuals as seen in Figure~\ref{sigma}.
We therefore conclude that the non-zero mean $\sigma$ for the line
residuals can be explained entirely by noise and is not due
to intrinsic variability of the star.

In comparing the variation of $\sigma$ for the
line residuals with the pulsation models  we can once again 
exclude all pulsations modes with $m$ $>2$ and
those with $k$ $\gg$ 10. 
Of the modes
having the smaller horizontal scaler, only an $m$ = 1 is consistent
with the apparent lack of line shape variations. An $m$ = 2 can
be hidden in the residual variations for an individual line, but it
seems to be excluded when using the residual variations averaged over
all lines.

\section{Discussion}
	
	Our analysis of  the spectral line shapes of $\tau$ Boo, both
through line bisectors and line residuals, indicate no variability
above the noise of the measurements. This
lack of spectral variability is consistent with
recent findings for this star by Brown et al. (1998).
If $\tau$ Boo is really a pulsating star,
then the only pulsation mode it could possibly have 
is an $m$ = 1, but with a low
horizontal scaler ($k$ $\approx$ 1), a value several orders of magnitude
less than the predicted value. This mode and  possibly 
an $m$ = 2 (although  less likely)
 could account for the observed RV variations in this star and yet
show no measurable changes in the spectral line shapes.

	One would think that these modes could be excluded
on the basis
of the low $k$ values, but this may not be the case in light of the
$k$-parameter problem discussed earlier. A calculation of
pulsations in a real stellar atmosphere along with the temperature
variations may result in a much lower ``observed'' $k$-value. 

	If the low $k$ values implied by this study do indeed represent
a true ratio of the horizontal to vertical velocities, then 
the pulsational amplitude 
can be used to estimate the photometric amplitude for the oscillations.
For both the $m$ = 1 and $m$ = 2 modes ($k$ = 1.2) the pulsational
amplitude is about 400 {\ms}. Buta \& Smith (1979) published expressions
for estimating the photometric amplitude due to geometrical effects.
These result in $\Delta$V $\approx$ 0.03 mag for the $m$ = 2 mode
and  $\Delta$V $\approx$ 0 mag for the $m$ = 1 mode.

	There may, of course, also be photometric variations due
to temperature effects and for very long periods these may dominate the
light curve (Buta \& Smith 1979). 
 The temperature variations can be estimated using
$$ |{{\delta t} \over {T}}| \sim \bigtriangledown_{ad} \sqrt{l(l+1)}
|{ \Omega \over \omega} {v_p \over v_e}|$$
where $\bigtriangledown_{ad}$ is the adiabatic temperature
gradient, $\Omega$ is the rotational frequency,
$\omega$ is the pulsational frequency, $v_p$ is the pulsational
amplitude and $v_e$ is the rotational velocity (Lee \& Saio 1990).
For $\tau$ Boo, $\omega$ $\approx$ $\Omega$, $v_p$ = 0.4 {\kms},
and $v_e$ = 15 {\kms}. For the $m$ (= $\ell$) = 1 mode $\delta$T
$\approx$ 170 K, and for the $m$ = 2 mode  $\delta$T $\approx$ 300 K. 
These variations result in $\Delta$$V$ $\approx$ 0.1 and 0.2 mag, 
respectively,
for the two modes, assuming that these temperature changes represent disk
integrated quantities. This is a reasonable assumption given the low
degree of the modes.
The lack of photometric variations seems to exclude such large
temperature
variations. However, as pointed out by Buta \& Smith, the temperature
variations  for very long period modes (i.e. much greater than the period
of the fundamental radial mode) are difficult to calculate and these
may have to include both non-adiabatic and non-linear effects. Consequently,
these approximate estimates, derived from linear theory, 
may not be appropriate for the hypothetical pulsation mode in $\tau$ Boo.
Furthermore, Buta \& Smith state that the light curve from 
temperature variations  may be
in-phase or out-of-phase with that due to the geometric variations.  
One could imagine
a scheme where the geometric variations nearly cancel those due to
temperature
variations producing a net zero change in the light level.  These 
may also conspire to produce small changes in the spectral line shapes, even
if $k$ is large. Clearly,
a detailed analysis of pulsations in a stellar atmosphere is required
to obtain better estimates of any predicted light curve. At face value,
however, the photometry does seem to exclude the possibility 
of such  low degree nonradial modes.

	Of course there is no reason to expect that the hypothetical
pulsation modes are spheroidal in nature. One possibility is
toroidal, or r-modes (Saio 1982). Because the atmospheric
motions for these modes are purely horizontal, they should
not be accompanied by photometric variations. It is not clear,
however, if these modes could produce appreciable RV variations.
For a star with intermediate inclination r-modes can 
produce spectral line distortions
(Unno et al. 1989), but these are not 
seen in our spectral data.

	Astrophysics, more than most sciences, has its level of
uncertainty. This is primarily because the objects of our study are
distant and cannot be scrutinized in laboratories, or up-close in
exquisite detail. The best we can do is collect high quality data and draw
the ``most probable'' hypothesis that is consistent with that data. There
will always be some probability, no matter how small, that the  conclusions
that are drawn 
are incorrect due to our ignorance of certain aspects
of natural phenomena. This is especially
true for extra-solar 
planet discoveries made with RV measurements since these
are indirect detections and some unknown process may be the cause of any
observed variability. One thing is certain,
if we had found spectral variability in the line shapes
of $\tau$ Boo with the RV period, then this would almost certainly have
killed the planet hypothesis.
Although our analysis shows no spectral variability
in $\tau$ Boo it does not prove beyond any doubt that this star
has a planetary companion. However, our measurements certainly add to the
considerable evidence weighing in favor of the planet hypothesis. It is
still the simplest and most likely explanation for the RV variability in this 
star.

Of the 5 short-period extra-solar planets,  two have been confirmed
using analyses of the spectral line shapes. Although it seems highly likely
that all the short-period  systems have  planets, it is
still worthwhile to search for any spectral variability in these other
stars.
We are in the process
of obtaining high resolution measurements for the other short-period
planets in order to confirm the planet hypotheses for these stars.

        This work was supported by NASA grant NAGW3990 and NAG5-4384.

\clearpage
\begin{deluxetable}{lcccccc}
\tablewidth{0pt}
\tablecaption{Journal of Observations}
\tablehead{\colhead{JD}  & \colhead{Phase} & \colhead{S/N} &
\colhead{JD}  & \colhead{Phase} & \colhead{S/N}}
\startdata
2450481.855 & 0.392 & 110 & 2450530.900 & 0.196    & 135 \nl
2450481.890 & 0.402 & 125 & 2450530.938 & 0.208    & 120 \nl
2450481.898 & 0.405 & 135 & 2450530.973 & 0.218    & 130 \nl
2450481.933 & 0.415    & 130 & 2450531.722 & 0.445    & 175 \nl
2450482.939 & 0.719    & 125 & 2450531.730 & 0.447    & 160 \nl
2450482.977 & 0.730    & 130 & 2450531.768 & 0.458    & 140 \nl
2450482.982 & 0.732    & 120 & 2450531.803 & 0.469    & 150 \nl
2450528.816 & 0.567    & 210 & 2450531.939 & 0.510    & 120 \nl
2450528.850 & 0.578    & 170 & 2450531.974 & 0.521    & 130 \nl
2450528.858 & 0.580    & 180 & 2450584.732 & 0.446    & 170 \nl
2450529.767 & 0.858    & 150 & 2450584.761 & 0.455    & 185 \nl
2450529.773 & 0.856    & 160 & 2450585.704 & 0.740    & 175 \nl
2450529.809 & 0.867    & 170 & 2450585.766 & 0.758    & 200 \nl
2450529.822 & 0.871    & 190 & 2450585.774 & 0.761    & 160 \nl
2450530.844 & 0.180    & 140 & 2450585.808 & 0.771    & 150 \nl
2450530.853 & 0.182    & 170 & 2450586.728 & 0.049    & 230 \nl
2450530.892 & 0.194    & 150 & 2450586.763 & 0.059    & 170 \nl
\enddata
\end{deluxetable}

\begin{deluxetable}{lcc}
\tablewidth{20pc}
\tablecaption{Spectral Lines for Bisector Measurements}
\tablehead{
\colhead{Wavelength}  &  \colhead{Species}        & \colhead{ $EW$} \\
\colhead{{\AA}}    & \colhead{} & \colhead{(m{\AA})}      }
\startdata
6213.3 & Fe I & 75   \nl
6336.9 & Fe I & 90 \nl
6439.1 & Ca I & 160 \nl
\enddata
\end{deluxetable}

\begin{deluxetable}{lcc}
\tablewidth{20pc}
\tablecaption{Spectral Lines for Residual Measurements}
\tablehead{
\colhead{Wavelength}  &  \colhead{Species}        & \colhead{ $EW$} \\
\colhead{{\AA}}    & \colhead{} & \colhead{(m{\AA})}      }
\startdata
6108.1 & Ni I & 55   \nl
6110.5 & ?    & 30  \nl
6113.3 & Fe II& 40    \nl
6116.2 & Ni I & 70  \nl
6224.0 & Ni I & 25  \nl
6335.4 & Fe I & 95  \nl
6336.9 & Fe I & 55 \nl
\enddata
\end{deluxetable}

\begin{deluxetable}{lcccc}
\tablewidth{25pc}
\tablecaption{Velocity Span Measurements }
\tablehead{
\colhead{JD}  & \colhead{Phase} & \colhead{$S_{6213}$} & 
\colhead{$S_{6337}$} & \colhead{$S_{6439}$}  \\
\colhead{}    & \colhead{} & \colhead{(m/s)} & 
\colhead{(m/s)} & \colhead{(m/s)} 
}
\startdata
2450481.894 & 0.404 &  113 & -12  & -108 \nl
2450482.966 & 0.727 & -158 &  70  & 170 \nl
2450528.841 & 0.575 &  222 & -139 & -61 \nl
2450529.793 & 0.863 & -191 & 300  & -146 \nl
2450530.863 & 0.185 & -97  & 1.2  & 30 \nl
2450530.937 & 0.207 & 18   & 64   & -55 \nl
2450531.756 & 0.455 & 198  &-191  & -60 \nl
2450531.957 & 0.516 & 170  & 105  & -175 \nl
2450584.747 & 0.451 & 38   & -125 & 71 \nl
2450585.747 & 0.753 & -70  & -41  & 204 \nl
2450586.746 & 0.054 & -129 & -29  & 131 \nl
\enddata
\label{spantable}
\end{deluxetable}

\pagestyle{empty}
\clearpage

\clearpage

\clearpage

%

\end{document}